\documentclass[aps,pra,floatfix,showpacs,twocolumn,superscriptaddress]{revtex4}
\usepackage{graphicx}
\usepackage{bm,amsmath,amssymb}
\usepackage{mathrsfs}
\usepackage{dcolumn}
\usepackage{mathrsfs}
\usepackage{cases}
\usepackage{color}
\usepackage{ulem}
\usepackage{indentfirst}
\usepackage{epstopdf}
\usepackage{grffile}

\DeclareGraphicsExtensions{.eps}

\newcommand{\be}{\begin{equation}}
\newcommand{\ee}{\end{equation}}
\newcommand{\bea}{\begin{eqnarray}}
\newcommand{\eea}{\end{eqnarray}}
\newcommand{\la}{\langle}
\newcommand{\ra}{\rangle}

\begin{document}

\title{Worm quantum Monte-Carlo study of phase diagram of extended
Jaynes-Cummings-Hubbard model}
\author{Huanhuan Wei}
\affiliation{College of Physics and Optoelectronics, Taiyuan University of Technology,
Shanxi 030024, China}
\author{Jie Zhang}
\thanks{Corresponding author:zhangjie01@tyut.edu.cn}
\affiliation{College of Physics and Optoelectronics, Taiyuan University of Technology,
Shanxi 030024, China}
\affiliation{State Key Laboratory of Quantum Optics and Quantum Optics Devices, Shanxi
Taiyuan 030024, Shanxi, China}

\author{Sebastian Greschner}
\affiliation{Institut f\"ur Theoretische Physik, Leibniz University Hannover, Appelstr.
2, DE-30167 Hannover, Germany}
\author{Tony C Scott}
\affiliation{Institut f\"ur Physikalische Chemie, RWTH-Aachen University, D-52056 Aachen, Germany}
\author{Wanzhou Zhang}
\thanks{Corresponding author:zhangwanzhou@tyut.edu.cn}
\affiliation{College of Physics and Optoelectronics, Taiyuan University of Technology,
Shanxi 030024, China}
\date{\today }

\begin{abstract}
Herein, we study the extended Jaynes-Cummings-Hubbard model mainly by
the large-scale worm quantum Monte-Carlo method to check whether or not a
light supersolid phase exists in various geometries, such as the one-dimensional
chain, square lattices or triangular lattices. To achieve our purpose,
the ground state phase diagrams are investigated. For the one-dimensional
chain and square lattices, a first-order
transition occurs between the superfluid phase and the solid phase and therefore there is no stable
supersolid phase existing in these geometries.
%which \tony{could/can?} be confirmed by
%density matrix renormalization group method simulation.
Interestingly, soliton/beats of the local densities arise if the chemical potential is adjusted in  the finite-size chain. However, this soliton-superfluid coexistence can not be considered as a supersolid in the thermodynamic limit.
Searching for a light supersolid, we also studied the Jaynes-Cummings-Hubbard model on triangular lattices, and the phase diagrams are obtained. Through measurement of the structural factor, momentum distribution and superfluid stiffness for various system sizes, a supersolid phase exists stably in the triangular lattices geometry and the regime of the supersolid phase is smaller than that of the mean field results. The light supersolid in the Jaynes-Cummings-Hubbard model is attractive because it has superreliance, which is absent in the pure Bose-Hubbard model. We believe the results in this paper could help search for new novel phases in cold-atom experiments.

\end{abstract}
\maketitle

%\begin{CJK*}{GBK}{song}

%%%%%%%%%%%%%%%%%%%%%%%%%%%%%%%%%%%%%%%%%%%%%%%%%%%%%%%%%%%%%%%%%%%%%%%%%%%%%%%%%%%%%%%%%%%%%%%%%%%%%%
%%%%%%%%%%%%%%%%%%%%%%%%%%%%%%%%%%%%%%%%%%%%%%%%%%%%%%%%%%%%%%%%%%%%%%%%%%%%%%%%%%%%%%%%%%%%%%%%%%%%%%
%%%%%%%%%%%%%%%%%%%%%%%%%%%%%%%%%%%%%%%%%%%%%%%%%%%%%%%%%%%%%%%%%%%%%%%%%%%%%%%%%%%%%%%%%%%%%%%%%%%%%%

%%%%%%%%%%%%%%%%%%%%%%%%%%%%%%%%%%%%%%%%%%%%%%%%%%%%%%%%%%%%%%%%%%%%%%%%%%%%%%%%%%%%%%%%%%%%%%%%%%%%%%
%%%%%%%%%%%%%%%%%%%%%%%%%%%%%%%%%%%%%%%%%%%%%%%%%%%%%%%%%%%%%%%%%%%%%%%%%%%%%%%%%%%%%%%%%%%%%%%%%%%%%%
%%%%%%%%%%%%%%%%%%%%%%%%%%%%%%%%%%%%%%%%%%%%%%%%%%%%%%%%%%%%%%%%%%%%%%%%%%%%%%%%%%%%%%%%%%%%%%%%%%%%%%
%%%%%%%%%%%%%%%%%%%%%%%%%%%%%%%%%%%%%%%%%%%%%%%%%%%%%%%%%%%%%%%%%%%%%%%%%%%%%%%%%%%%%%%%%%%%%%%%%%%%%%
%%%%%%%%%%%%%%%%%%%%%%%%%%%%%%%%%%%%%%%%%%%%%%%%%%%%%%%%%%%%%%%%%%%%%%%%%%%%%%%%%%%%%%%%%%%%%%%%%%%%%%
%%%%%%%%%%%%%%%%%%%%%%%%%%%%%%%%%%%%%%%%%%%%%%%%%%%%%%%%%%%%%%%%%%%%%%%%%%%%%%%%%%%%%%%%

\section{introduction}

A supersolid $(SS)$ phase is a novel quantum state with is both superfluid and solid\cite{condensation,sov,speculations,BEC3}. In the past ten years, with the great progress of quantum
manipulation of atomic technology, scientists have tried to observe the $SS$ phase
in ultra-cold atomic systems. A breakthrough occurred in 2017 when
two groups observed the $SS$ phase in cold atoms\cite{supersolid formation,supersolid
properties}. In 2019, three independent groups found the $SS$ phase experimentally\cite{dipolar quantum
gas,DipolarQuantum Droplets,supersolid behaviors}.

Physical models include the extended Bose-Hubbard (BH) model\cite{Supersolids
versus Phase Separation}, the two-component BH model\cite{2CM1,2CM2,2CM3}, the paired BH model\cite{liangP1,guoP2,wangP3,chenP4,troyP5,guoP6},
the Bose-Fermi mixture system\cite{BF1,BF2}, the Fermion
system\cite{fm1} and the spin
system\cite{2spin1,spin2}, the $SS$ phase in
spin-orbit coupled systems\cite{Ultracold Atomic Bose Condensates}, the BH model with next-nearest neighborhood interactions \cite{next},
and dipolar bosons\cite{5}.

The $SS$ model mentioned earlier does not involve the quantum optical
cavity model and its extensions. Most quantum optics models based on
single-cavity systems have interesting phases, such as the super-radiation
solid phase in a single-cavity system coupled with an optical lattice \cite{A
Superradiant Solid}, that is, coexistence of a super-radiation order and
an atomic solid order.
% TCS: no paragraph should start with However
However, here we will extend the single-cavity case to the lattices 
of multiple cavities
with various geometries, i.e. one-dimensional lattices,  square lattices, and triangular lattices.
The Jaynes-Cummings models sitting in each lattice site form
the so-called Jaynes-Cummings-Hubbard (JCH)\cite{QED,SMI} model with additional photons hopping between cavities.

Experimentally, the JCH model can be realized by a coupled-transmission-line resonator\cite{Low} or trapped ions\cite{Exp}. Moreover, theoretical studies\cite{MF,MF1, GL, strong} and reliable numerical methods, such as the density matrix renormalization group
algorithm(DMRG)\cite{AM,Fermionized} and the quantum Monte-Carlo(QMC)\cite{Dynamical critical,zhaojize}
 method are also investigated. Furthermore, various 
topics including  fractional
quantum Hall Physics\cite{Fractional}, quantum transport\cite{Quantum transport},
quantum-state transmission\cite{Quantum-state}, on-site disorder\cite{MF1,Disorder} and phase transitions\cite{MF,MF1} are explored intensively.

In regards to the anti-ferromagnetic correlation between the excitations of each atom, 
the JCH model
can be considered as an extended JCH model.
There are still interesting questions to be discussed for the extended JCH models.
%The authors of these  papers have
%, and explore the scope and mechanism of the existence of
%super-radiation (light) super-solid. Super-radiation super-solid, a novel
%quantum phase in which three orders of super-radiation, super-fluid and
%solid phase coexist\cite{guolijuan}.
\renewcommand{\theenumi}{\roman{enumi}}%
\begin{enumerate}
\item
For pure hard-core BH models, there is no $SS$ phase in the bipartite lattices.
It is still not clear whether or not there is a $SS$ phase in the JCH model 
at the bipartite lattices,
as is the case for JCH models on one-dimensional lattices and square lattices.
Meanwhile, Ref.~\cite{light-ss} shows the $SS$ phase with nearest interactions between cavities
by using the cluster mean field method. They also expect a reliable method to 
vindicate their findings.
\item
In previous work \cite{guolijuan}, we used the DMRG method\cite{DMRG1,DMRG2} to give the paired $SS$ phase 
for the JCH model on different ladders. 
Furthermore, for the JCH system on the two-dimensional triangular
lattice, we presented the mean field (MF) result to give a preliminary
phase diagram using the cluster mean field method. Although the cluster size
has up to 36 sites, we still need more reliable methods to confirm the stability
of the $SS$ phase.
\end{enumerate}

To clarify the above questions, we mainly use the large-scale 
worm algorithm of the quantum Monte-Carlo (wQMC) method\cite{wqmc1,wqmc2,wqmc3,wqmc4}, to
simulate the model on various geometries.% The reason we choose the method is that
%the method needs less computer memories than DMRG method.
The global
phase diagram by the MF method is plotted as background.
%This paper also intends to use a new machine learning method, combined with
%the extended configuration given by the worm Quantum Monte Carlo (QMC)
%Method, to study the super-radiation (light) super-solid phase, and use the
%snake model method to give a transcendental mean field phase diagram.

We find that, for bipartite lattices,
there is no light $SS$ phase  in the 
extended JCH model. 
The previously predicted $SS$ phase by the MF method\cite{light-ss} is not
stable, i.e., the transition between the solid phase to the superfluid phase is of first order
with obvious jumping of the order parameters such as the superfluid stiffness and 
structural factors.
To search for the light $SS$ phase, the JCH model on the two-dimensional triangular lattices 
is studied and a stable light $SS$ phase is found.

The outline of this paper is as follows:
Sec.~\ref{sec:method} introduces the model, methods and the measured quantities.
Sec.~\ref{sec:bi} shows the results on bipartite lattices such as 1D and 2D square lattices.
Sec.~\ref{sec:tri} gives the results of the JCH models on triangular lattices. 
Conclusions are made at the end in Sec.~\ref{sec:con}.

\section{model, methods and the measured quantities}
\label{sec:method}
\subsection{Model and its mapping}
%{\ In Figs.~\ref{h}(a) and (b) show the JCH model on one-dimensional and triangular cavity lattices. For convenience, we decompose each unit cell into two (three) sublattices labeled by A and B (A, B and C). As shown {\ In Fig.~\ref{h}(c), $a_i\sigma_i^\dagger$ means that a photon is absorbed and an atom excitation forms simultaneously. }}

\begin{figure}[t]
\includegraphics[width=0.4\textwidth]{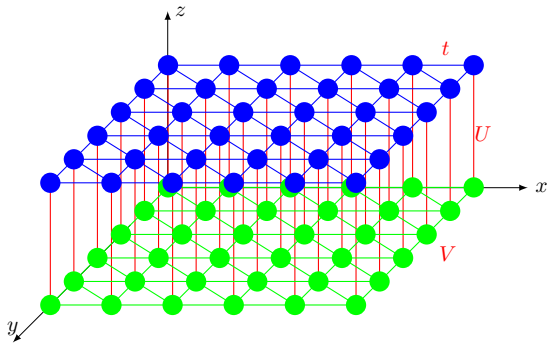}
\caption{(Color online) The mapped two-layer triangular lattices for the JCH model, where the top and
bottom layers are denoted as photon and atom layers respectively. The
hopping of the photons $t$, atom-photon coupling $U$, and interactions between atom excitations $V$ are labeled. A one-dimensional chain and a square lattice are illustrated in appendix A.}
\label{fig:latt}
\end{figure}

The extended JCH model includes a nearest repulsion between excitations of 
the cavities with strength  $V$, which is
different from the atom-atom repulsion 
interaction. The model has many cavities and 
each cavity can be considered a site in the lattices. In each cavity, there is a 
two-level atom. Simultaneously, photon tunneling between cavities is considered.

To simulate the model conveniently, the model could be
mapped unto a pure BH model on two layered geometries, e.g  two-layer triangular lattices 
as shown in Fig.\ref{fig:latt}, where
the top layer and bottom layer describe the state of the photon and the level of the atom,
respectively. For a specific site $\textbf{r}$=($x$, $y$), if the two-level atom sits
in the ground state, then the state at ($x$, $y$, $z=1$) should be empty and
the excited state, occupied. Similarly,
the state at ($x$, $y$, $z=2$) describes the
photon number in each cavity.
%TCS is the photon number 2?  

In comparison to the pure BH model, the interaction and hopping between layers 
are different.
In the photon layer, only photon hopping without repulsion exists and in the
atom layer, only repulsion without the hopping term exists.
%
%The model could be considered as a hardcore BH model if only at most one photo is allowed
%in each site on the upper layer.
The extended JCH Hamiltonian is defined as
\begin{equation}
\begin{aligned} H=&\sum_{i}(H_i^{JC}-\mu n_i)-t\sum_{\la i,j\ra}(a_i^\dagger
a_j+\rm H.c.)\\ &+\sum_{\la i,j\ra} V n_i^\sigma n_j^\sigma, \label{a}
\end{aligned}
\end{equation}%
where the total number of excitations is $\rho \equiv \sum_{i}{n}%
_{i}=\sum_{i}(n_{i}^{\sigma }+n_{i}^{a})$, $\mu $ is the chemical potential,
$%
a_{i}^{\dagger }$ and $a_{i}$ are respectively the photon creation and annihilation
operators at lattice site $i$, and the term $(a_i^\dagger
a_j)$ with strength $t$  represents the photon hopping between cavities.
%
%On each cavity site $i$, the two-level atom is contained \cite{Dynamical
%critical}.
The on-site coupling between the photons and the atom on each
site $i$ can be described by the JC Hamiltonian $H_{i}^{JC}$
\begin{equation}
\begin{aligned} H_i^{JC}=&\omega n_i^a+\varepsilon n_i^\sigma+U (a_i^\dagger
\sigma_i+a_i \sigma_i^\dagger), \label{b} \end{aligned}
\end{equation}%
where $U$ is the
atom-photon coupling strength and represents the tunneling between 
layers, $\omega $ is the frequency of the model of the photon creation and
annihilation operators at lattice site $i$, and, $\varepsilon $ is the transition
frequency between two energy levels. For simplicity, we restrict $\omega = \varepsilon=0$. Operators $n_{i}^{a}=a_{i}^{\dagger }a_{i}$ and $%
n_{i}^{\sigma }=\sigma _{i}^{\dagger }\sigma _{i}$ are the photon number and
the number of excitations of the atomic levels, respectively.  The Pauli matrices $\sigma _{i}^{\dagger }$ ($%
\sigma _{i}$) represent the raising (lowering) operators for the atom levels.

\subsection{Methods and the quantities measured}

The conclusive results are mainly obtained by the wQMC method, and the details 
can be seen in Refs.~\cite{wqmc1, wqmc2,wqmc3, wqmc4}. Here,
in the real simulations, the inverse temperature  $\beta=1/k_bT$
takes larger values, namely, $\beta=100,~500,~\cdots,~1500$ for allowing
the systems to converge to the ground state.
%
%%The difficulty in clarify the $SS$ phase.
%why (?)\\
%i)We have simulated  the global phases diagram by mean field methods and find
%possible signature of supersolid around the tips of solid phase. However, it is
%well-known that MF method's results need to be verified by other more reliable method
%as MF ignore quantum fluctuations.\\
%ii)DMRG method is very suitable to simulate (quasi) one-dimensional JCH model. However
%for two-dimensional problems, it is still a challenge, especially for the regimes  where energy gap is small.
% \\
We have tried a stochastic series expansion QMC method\cite{expansion,directed loops}. However, due to local atom-photon coupling
and the no-hopping term between each atom layer, the updating efficiency
is very slow in low temperature regimes.
The needed quantities for the worm QMC algorithm are:\\
\renewcommand{\theenumi}{\Roman{enumi}}%
\begin{enumerate}
\item Local photon density $\rho_i^a=\left\langle n_i^a\right\rangle$ and local atom excitation  $\rho_i^{\sigma}=\left\langle n_i^{\sigma}\right\rangle$.\\
\item Superfluid stiffness\cite{stiff}
\begin{equation}
\rho _{s}=\sum_{r=1}^d\frac{L^{2-d}\left\langle W_{r}\right\rangle ^{2}}{2d\beta t},
\end{equation}%
where $W_{r}$ is the winding number of the photons in the upper layer in the
$x$ or $y$ direction.
The stiffness characterizes the non-diagonal long range order of the system.
\item Structural factor of photons, given by
\begin{equation}
S(\mathbf{Q})/N=\left\langle \rho_\mathbf{Q}\rho_\mathbf{Q}^{\dagger}\right\rangle
\end{equation}
where $\rho_\mathbf{Q}=(1/N)\sum_{i}n_{i}^a\exp(i\mathbf{Q}\mathbf{r}_i)$, $N=L\times L$ is the total number of physical sites for 2D systems and $N=L$ for 1D systems.
For a one-dimensional lattice, the wave vector is at $%
\mathbf{Q} =(\pi,0)$. In real space, the density of excitation obeys configurations
of the form ($101010\cdots$) or ($010101\cdots$). The phase here is called the solid I $(SI)$ phase. For square lattices, the solid with $\mathbf{Q} =(\pi,\pi)$ is also called the $SI$ phase.

For the two-dimensional triangular lattices, the wave vector is at $\mathbf{Q} =(4\pi/3,0)$ and excitation densities obey configurations of the form ($001001\cdots$) or ($110110\cdots$). The densities 
are $1/3$ and $2/3$, and called the $SII$ phase and $SIII$ phase, respectively.
\item
%Green function
%\be G(r)=a_0^{\dagger} a_r\ee
%and momentum distribution
Momentum distribution given by
\be
n\left( \textbf{k}\right) =\frac{1}{N}\sum\limits_{j,j^{^{\prime }}}\left\langle
a_{j}^{\dagger }a_{j^{^{\prime }}}\right\rangle e^{  i\textbf{k}\left(
\textbf{r}_j-\textbf{r}_{j^{^{\prime }}}\right) }
\label{eq:nk}
\ee
In the $SS$ phase, $S(\mathbf{Q})/N$ and $\rho _{s}^{x(y)}$ are both nonzero in the thermodynamic limit.
Moreover, the global phase diagram is plotted with $\Psi=\la a\ra$ in the MF frame, which is illustrated in appendix C for completeness.

The above quantities of $\rho$, $S(\pi)/L$ and energy density $E$ are calculated by exact diagonalization and wQMC methods.The results
are consistent with each other and shown in appendix B.
\end{enumerate}

\section{Results for the JCH model on bipartite lattices }
\label{sec:bi}

In this section, we focus on the results of the JCH models on both the 1D and square lattices.

\begin{figure}[t]

\includegraphics[width=0.45\textwidth]{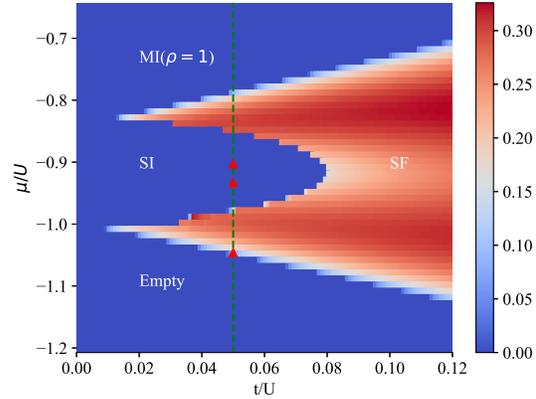}
\caption{ (Color online) Phase diagram and the detailed description of $\Psi$ for
the 1D hardcore extended JCH model with $V/U=0.4$ by the MF method. The green dashed line labels the position scanning by wQMC and
the red triangular symbols are the phase boundaries given by the wQMC method.}
\label{hdph}
\end{figure}

\begin{figure}[tb]
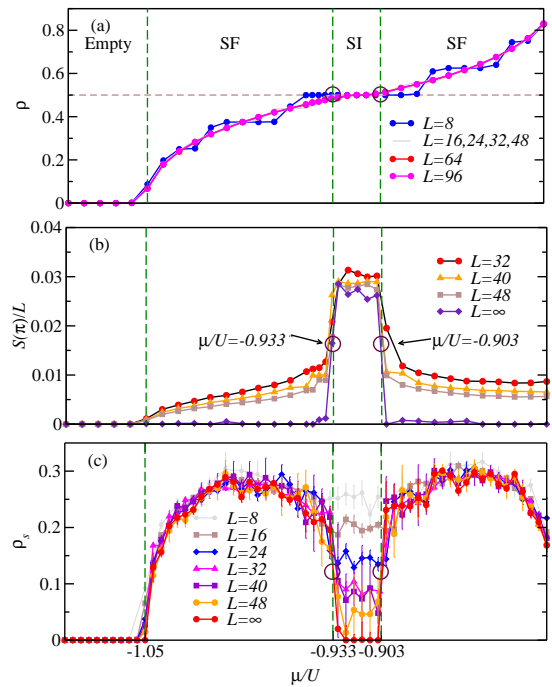

\includegraphics[width=0.395\textwidth ]{fig3_a_1dn.eps}\vskip 0.1cm %
\includegraphics[width=0.4\textwidth ]{fig3_b_1dsqe.eps} \vskip 0.1cm %
\includegraphics[width=0.4\textwidth ]{fig3_c_1drs1.eps} %
\caption{(Color online) Monte Carlo results of (a) excitation densities $\rho$ (up to $L=96$), (b)
structural factors $S(\pi)/L$ and (c) superfluid stiffness $\rho_s$ as a function of $\mu%
/U $ with system sizes $L =8,~16,~24,~32,~40,~48$ and $\infty$ at parameters $V/U=0.4$ and $t/U=0.05$.
 }
\label{hdmc}
\end{figure}

\subsection{Hardcore 1D JCH model}
%TCS: Two labels for same subsection? - I nulled out one of them
\label{sec:1d}
%\label{subsec:hardcore} 
For the 1D hardcore extended JCH model, the maximum number of
photons is restricted to be one in each cavity. It is well known that
the number of photons is not fixed in a grand canonical ensemble\cite{Dynamical critical}. Therefore, the softcore photon system has to be checked and the Physics does not change\cite{guolijuan}.

Fig.~\ref{hdph} shows the MF phase diagram\cite{opt,cluster2,cluster3,cluster4,cluster5},
which contains the empty, $SI$, $SS$, $SF$ and $MI$($\rho=1$) phases, by plotting $%
\Psi$ in the plane ($t/U$, $\mu/U$).
%Here very narrow regimes around the $SI$ tips with signature $\Delta \Psi \neq 0$  (now shown).
% in the tips around
%the $SI$ phase which means that there are possible supersolid phase there.

%The phase diagram is not exactly symmetric with particle-hole symmetry at $\mu/U=-0.915$, which is a bit different from the case of hardcore bosons on bipartite lattices~\cite{LIUJIAO,zhengfang1,zhengfang2} as a result of the atom-photon coupling.

As $t/U$ is small, and $\mu/U\textless-1$, the system is in an empty phase
with $\rho=0$ and $\Psi=0$. The system sits in the $MI$($\rho=1$) phase
if $\mu/U>-0.83$. Moreover, the
$SI$ phase appears between the two regimes. As $t/U$ gets larger, the system enters the $SF$ phase. As
discussed in Ref.~\cite{SMI}, for a large hopping $t/U$, the ground state
energy becomes negative and can be made arbitrarily small by increasing the
total number of excitations. Here, since the maximum number of photons is
fixed, the $SF$ remains stable.

%(b) The $SS$ order $\Delta \Psi$ vs $t/\protect\beta$
%with $\protect\mu/\protect\beta$ from $-0.99$ to $-0.95$.

%{\ In Fig.~\ref{hdmc}(a), we scan $t/U$ along $\mu/U$ from $-0.95$ to $-0.99$,
%$\Delta \Psi$ is obviously nonzero. The $SS$ phase ($\Delta\Psi \ne 0$)
%emerges around the tips of the $SI$ phase. A similar phase diagram of the JCH
%model has been obtained on the square lattice in Ref.~\cite{Supersolid}.

The question about whether or not the $SS$ phase could exist in the thermodynamic
limit can now be verified by the wQMC method.
Fig.~\ref{hdmc}(a) shows the results of excitation densities as a function of $\mu/U$ with system sizes $L =
8,~16,~24,~32,~40,~48,~64,~96$ and $\infty$. %This picture includes dimensions ns(1),ns(2),ns(3) of 96,2, 1; temperature $\beta=1/T(500)$; It
%can be seen from the figure that
If $\mu/U$ increases, the excitation
density increases.  When $\mu/U$ increases to about $-0.9$, a plateau of density ($\rho=0.5$) appears.

Fig.~\ref{hdmc}(b) shows the structural factor $S(\pi)/L$ of photons obtained as a function of $\mu/U$ with sizes $L =32,~40,~48$ and $\infty$ with 
% TCS: no need to use "as" twice.  Once is enough
the sufficiently low temperature $\beta=1500$.
 % The results of $\beta=500$ are also calculated for the purpose of check (not shown) here.

Consistent with the excitation density, nonzero $S(\pi)/L$ has a platform
in the interval $-0.933\textless \mu/U\textless -0.903$, and  
simultaneously, superfluid stiffness $\rho _s^x$ becomes zero in the thermodynamic limit in Fig.~\ref{hdmc}(c).
% TCS: zeros should be zero
The two signatures clearly demonstrate that a $SI$ phase exists in the regimes.

By doping vacancies or excitations on the $SI$ phase,
$S(\pi)/L$ converges to zero and $\rho _s^x$ becomes non-zero
at $L \rightarrow \infty$. The behavior of the jump at those two ends of the platform
represents clear first-order transitions between the $SF$ and $SI$ phases, and
% TCS: there seems to be many a transition for each L case.
obviously no $SS$ phase exists. The $SS$ phase is unstable against the phase separation\cite{separation,separation2,separation3,BF3}.

Here, we just show the quantities sampled from the upper layer.
Moreover, the phase transition points from the empty phase to the $SF$ phase, 
the predictions by the MF and wQMC methods being completely the same at $\mu/U=-1.05$.

\begin{figure}[htb]
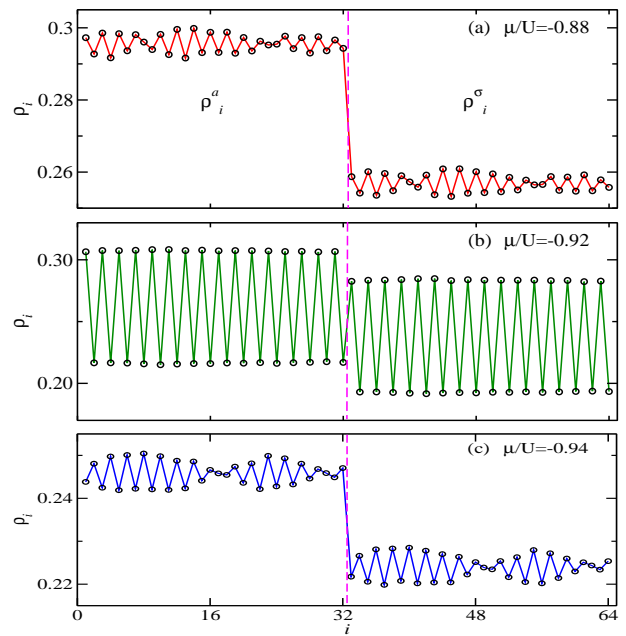

\includegraphics[width=0.45\textwidth,height=0.15\textwidth]{fig4_abc_pa.eps} \vskip 0.1cm %
\includegraphics[width=0.45\textwidth,height=0.15\textwidth]{fig4_abc_pb.eps} \vskip 0.1cm
\includegraphics[width=0.45\textwidth,height=0.15\textwidth]{fig4_abc_pc.eps}%
\caption{(Color online) Local photon density $\rho^{a}_i$, local atom excitation density $\rho^{\sigma}_i$ and solitonic signature in upper and bottom layers as a function of position $i$ with $V/U$=0.4 and $t/U=0.05$. (a)$\mu/U$=-0.88, with on-site excitation density $\rho>1/2$. (b) $\mu/U$=-0.92, with $\rho = 1/2$. (c) $\mu/U$=-0.94, with $\rho<1/2$.}
\label{cns}
\end{figure}

In the regime $-1.05\textless \mu/U\textless -0.933$, $S(\pi)/L$ for a finite system size is not zero (see Fig.~\ref{hdmc}(b)). To further understand $S(\pi)/L$ , we illustrate the local photon densities $\rho^{a}_i$ and the local atom excitation density $\rho^{\sigma}_i$ function of position $i$ under different values of chemical potential $\mu$ in Fig.~\ref{cns}.
The commensurate density distribution arises in Fig.~\ref{cns}(b), which confirms the periodic ground state long-range crystalline order in the solid phase. The on-site excitation density in the solid phase is calculated to be $\rho = 1/2$, which includes both photons and atoms. Therefore e.g. the local photon density $\rho^{a}_i$ in the upper layer oscillates between $0.19$ and $0.31$ with the average value being $0.25$.

In Figs.~\ref{cns}(a) and (c), the beats or soliton patterns\cite{6,beats} arise, which means changing the chemical potential $\mu$ (or removal/addition of the photons) will give rise to the soliton patterns. Only the uniform density oscillation is seen but no beats or soliton patterns appear in Fig.~\ref{cns}(b).
The number of solitons will increase as the photons are further added or removed, meanwhile the density oscillation becomes weak.
In other words, starting with the $SI$ phase, and changing the chemical potential $\mu$ leads to a solitons+$SF$ crossover instead of the $SS$ phase. In the thermodynamic limit, the crossover region will vanish, and the system will experience first-order transitions immediately.

\begin{figure}[htb]
\includegraphics[width=0.45\textwidth,height=0.225\textwidth]{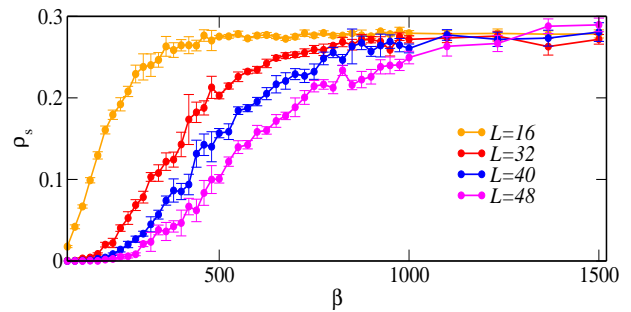}
\caption{(Color online) Simulation results by the wQMC method of superfluid stiffness $\rho_s$ as a function of $\protect\beta$ for the hardcore 1D JCH model. The parameters are $V/U=0.4$, $\mu/U=-1$, and $t/U=0.05$, and the system sizes are $L=16,~32,~40$ and $48$, respectively.}
\label{beta}
\end{figure}

Usually, for the BH model, the ground state will be reached if $\beta t >L$.
For the JCH model, through careful checking, $\beta$ should be much larger than $L/t$.
In particular, for $L=48$, 
$\rho_s$ should reach convergence here at $\beta =48/ 0.05 \approx 1000$.
% TCS: the author never really addressed my question: so I am making a guess.
 However, as shown only in  Fig.~\ref{beta} with $L=16,~32,~40$ and $48$, the temperature should be sufficiently low enough for a ground state. Therefore, in the section, $\beta$ is chosen as 1500.

\subsection{ Hard core JCH model on square lattice}
\begin{figure}[tb]
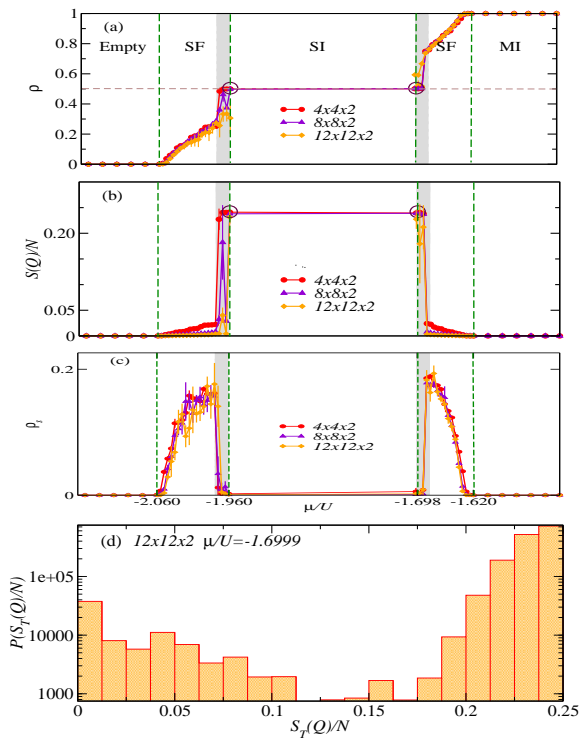

\includegraphics[width=0.395\textwidth,height=0.12\textwidth ]{fig6_a_squarenn.eps}\vskip 0.1cm %
\includegraphics[width=0.4\textwidth,height=0.12\textwidth ]{fig6_b_squaresqe.eps} \vskip 0.1cm %
\includegraphics[width=0.4\textwidth,height=0.12\textwidth ]{fig6_c_squarers.eps}  \vskip 0.1cm %
\includegraphics[width=0.423\textwidth,height=0.16\textwidth ]{fig6_d_squaresqb1.eps} \vskip 0.1cm %
\caption{(Color online) Simulation results by the wQMC method of (a) $\rho$ and
(b) $S(Q)/N$ (c) $\rho_s$ as a function of $\mu/U$ for the JCH
model on the square lattices. The parameters are $t/U=0.025$, $\beta=1000$ with  $L=4,~8,~12$, respectively.
Dashed regimes are for first-order transitions. (d) Histogram of $S_T(Q)/N$ obtained at the phase transition point $\mu/U=-1.6999$ for the system size $L=12$.}
\label{fig:2dh}
\end{figure}

Although the square lattice is a bipartite lattice, the Physics between 
the 1D and 2D models may be different.
We still need to perform powerful wQMC simulations on the 2D geometries.
Various global phase diagrams in the MF frame have been shown in Ref.~\cite{light-ss}.
The purpose here is to check the results by the reliable wQMC method. As shown in Fig.~\ref{fig:2dh}, we also set $t$ at the fixed value $t/U=0.025$ and then measured
$\rho$, $S(Q)/N$, and $\rho_s$ as a function of $\mu/U$ for the JCH
model on the square lattices. The inverse temperature is $\beta=1000$ and system sizes are $L=4,~8,~12$ and $16$, respectively.

In a manner similar to the 1D model, we still see the $SI$ phase (0,1,0,1) order in both directions and the wave vector is $\textbf{Q}$=($\pi$, $\pi$) in the range of $-1.960\textless \mu/U\textless -1.698$, with signatures
$\rho=0.5$, $S(Q)/N\ne 0$, and $\rho_s=0$. At the two ends of the $SI$ phase, $S(Q)/N$ and $\rho_s$ change discontinuously, which clearly indicates that no $SS$ phase exists in the square lattices. The histogram of $S_T(Q)/N$ obtained at the phase transition point in Fig.~\ref{fig:2dh}(d) demonstrates the two peaks which also indicate the first-order transition between the $SI$ and $SF$ phases. Here, the total structure factor $S_T(Q)/N$ is defined by replacing $n_{i}^a$ with $n_{i}^{\sigma}+n_{i}^a$, as a signature of the double peaks is more clear.
% Are the double peaks the two peaks mentioned earlier?
In addition, we observe empty, $SF$ and $MI$($\rho=1$) phases as expected. The critical points of the empty-$SF$ phase transition is -2.06, and similarly for the results from the MF methods.

\section{Results for the JCH models on triangular lattices}
\label{sec:tri}
\begin{figure}[htb]
\includegraphics[width=0.5\textwidth]{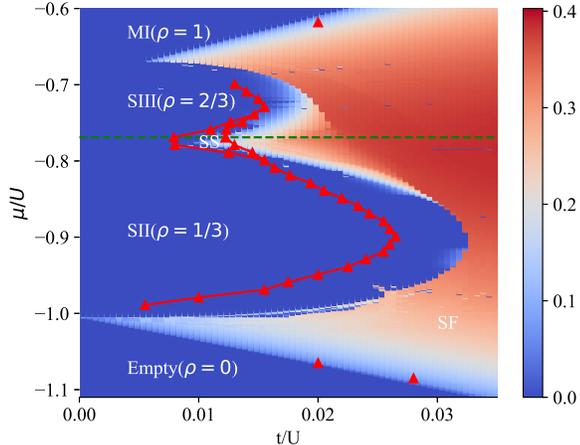}
\caption{The triangular lattices phase diagram and the detailed description of $\Psi$ for 2D JCH model with $V/U=0.4$ by the MF method. The red triangular symbols are the phase boundaries given by the wQMC method, and the green dashed line labels the cut scanning with $\mu/U=-0.77$.}
\label{xt}
\end{figure}

The BH model on the triangular lattices has been studied extensively\cite{tri2,tri3,zhangP7,Mishra15,zhangT1,zhangT2,zhangT3},
 Fig.~\ref{xt} shows the phase diagram obtained from the MF method,
which contains the empty, $SII$, $SIII$, $MI$($\rho=1$), $SS$ and $SF$ phases in the plane ($t/U$, $\mu/U$).
The colored symbols denote the numerical results obtained by wQMC methods.
The phase diagram of the pure BH model on the triangular lattices has been obtained\cite{tri2,tri3}, and is symmetric
about the particle-hole symmetric point $\mu/V=3$. Here, for the JCH model, the phase diagram is not symmetric due to
the atom-photon coupling, and a similar ``symmetric  point'' 
can be found between the $SII$ and $SIII$ phases, which locates itself at about $\mu/U=-0.775$ (green dashed line) through the wQMC method.

\begin{figure}[htb]
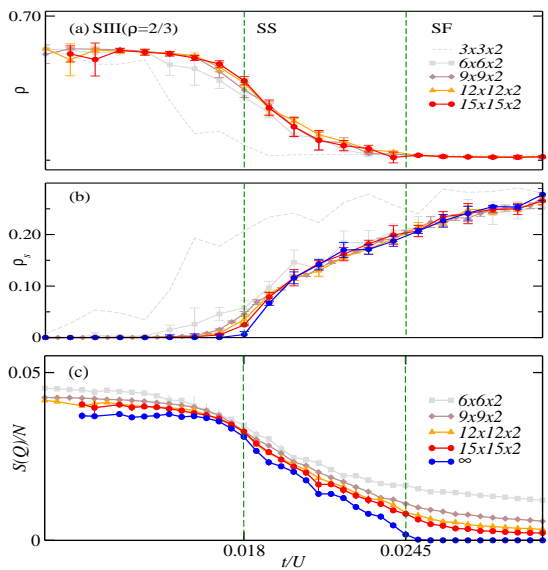

\includegraphics[width=0.4\textwidth,height=0.12\textwidth]{fig8_a_tritn.eps} \vskip 0.1cm
\includegraphics[width=0.4\textwidth, height=0.12 \textwidth]{fig8_b_tritrs.eps} \vskip 0.1cm
\includegraphics[width=0.4\textwidth, height=0.16 \textwidth]{fig8_c_tritsqe.eps}
\caption{
Simulation results by the wQMC method of (a) $\rho$,
(b) $S(Q)/N$,  and (c) $\rho_s$ as a function of $t/U$ for the JCH
model on the triangular lattices. The parameters are $\protect\mu/U=-0.77$, $
\protect\beta=500$ and the system sizes are $L\times L\times 2$, where $L=6,~9,~12$ and $15$, respectively.}
\label{trimc}
\end{figure}

The system is in an empty phase when $\mu/U$ is very small (bottom of the phase diagram) and in
the $MI$($\rho=1$) phase when $\mu/U$ is large (top of the phase diagram).
In the middle regimes, around $\mu/U=-0.775$, the $SII$ ($\rho=1/3$) and  $SIII$ ($\rho=2/3$) phases emerge.
As $t/U$ gradually increases, the system enters into the $SF$ phase. 
In a manner similar to the pure BH model, the phase transitions between them are as follows: the solid-$SF$ transition is first order, and the $SF$-$MI$ transitions are continuous.

Surrounded by the two types of solid phases and the $SF$ phase, the $SS$ phase emerges in the closed regime illustrated by lines with triangular symbols as shown in Fig.~\ref{xt}.

%Contrast to BH model\cite{tri2,tri3} or pair BH model\cite{wangP3,zhangP7},  we have not observed obvious jumps of the quantities such as $S(Q)/N$ or $\rho_s$.

Fig.~\ref{trimc} shows the wQMC simulations of $\rho$, $\rho _{s}$, $S(Q)/N$ as a function of $t/U$. The parameters are $\mu /U=-0.77$, inverse
temperature $\beta =500$ and the mapped system sizes are $L\times L\times 2$
, where $L=3,~6,~9,~12,~15$ and $\infty$, respectively. In the regimes of $t/U\textless 0.018$, the systems are trapped in the $SIII$ ($\rho=2/3$)
solid phase, with signature of $S(Q)/N \ne 0 $ and $\rho _{s}=0$. When $t/U$ increases, induced vacancies lead to the decrease from  the excitation density and both $S(Q)/N$ and $\rho _{s}$ are nonzero. In particular, in the region of $0.018\leq t/U\leq 0.0245$, the system sits in the $SS$ phase stably in some special parameter regime of the triangular JCH model.

\begin{figure}[tbh]
\includegraphics[width=0.23\textwidth]{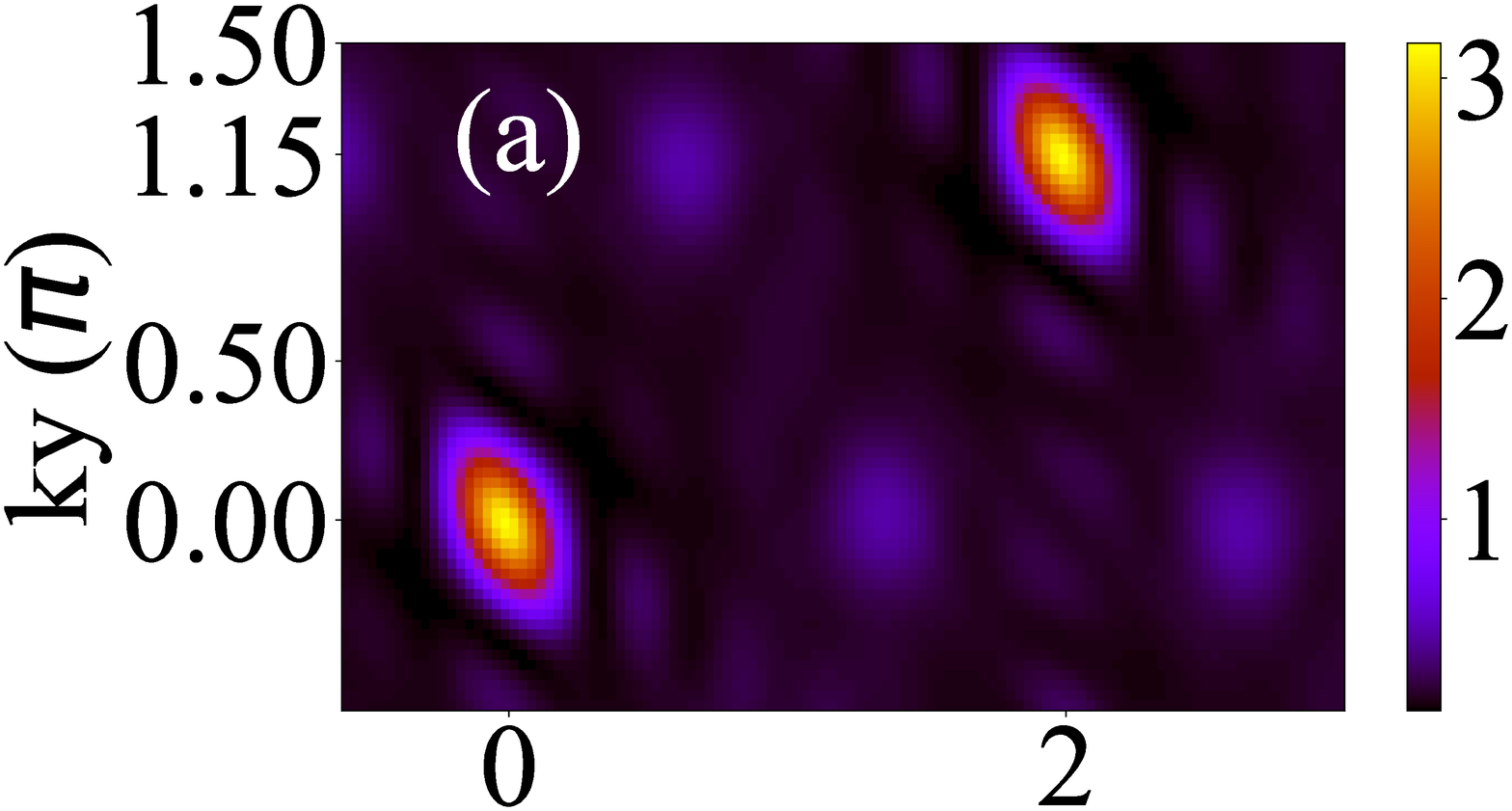} %
\includegraphics[width=0.23\textwidth]{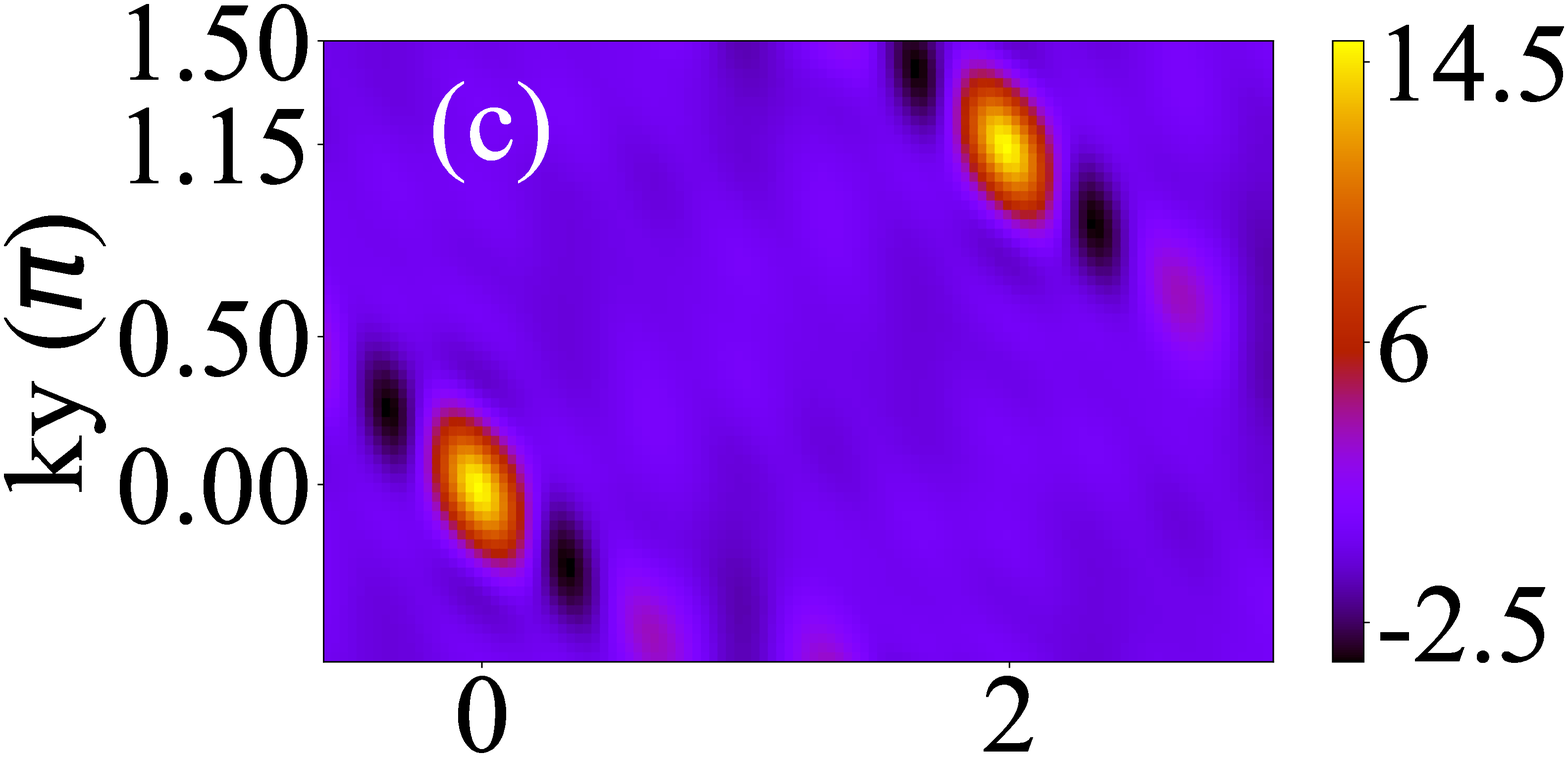} \vskip 0.1cm
\includegraphics[width=0.23\textwidth]{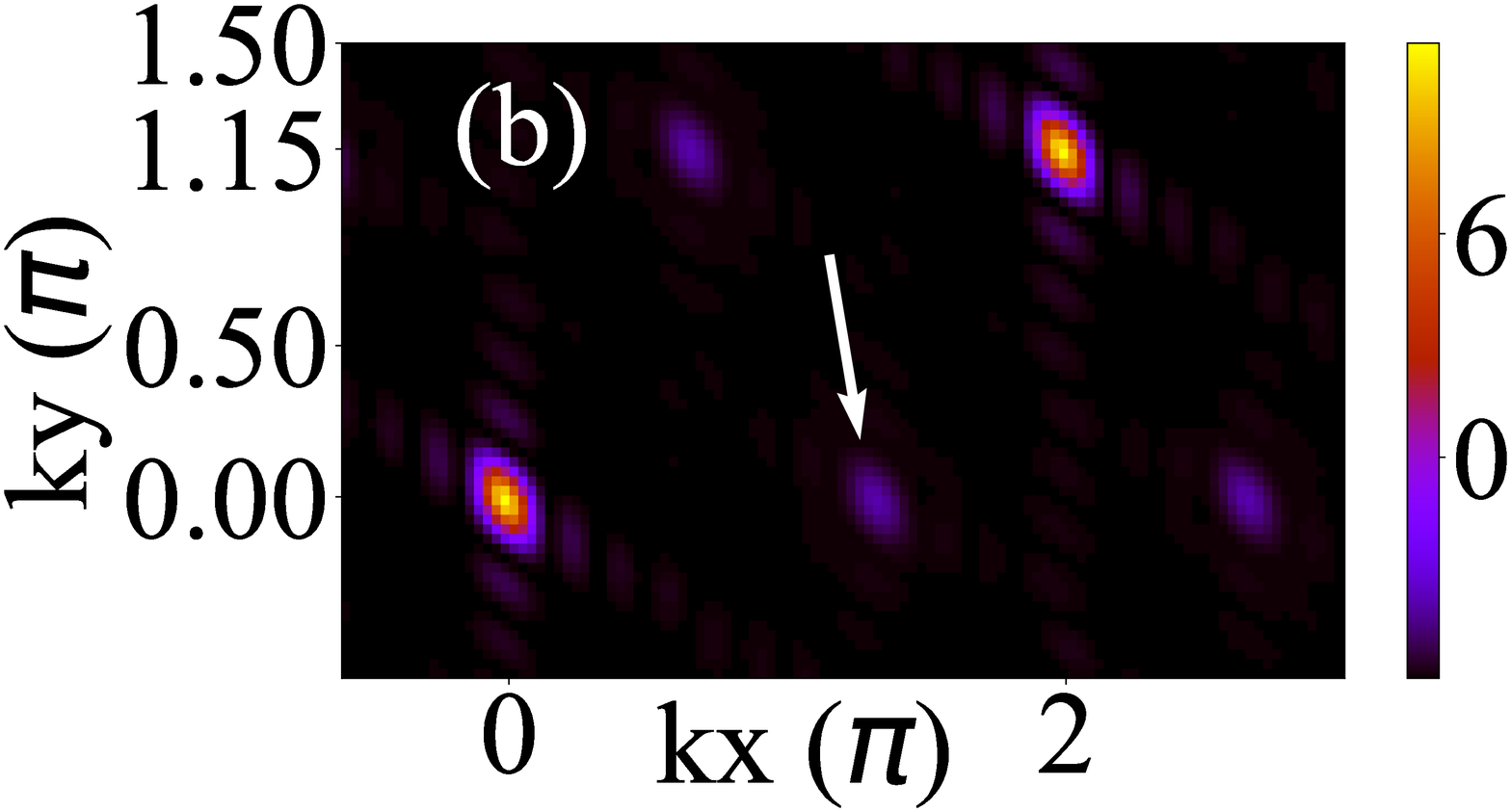} %
\includegraphics[width=0.23\textwidth]{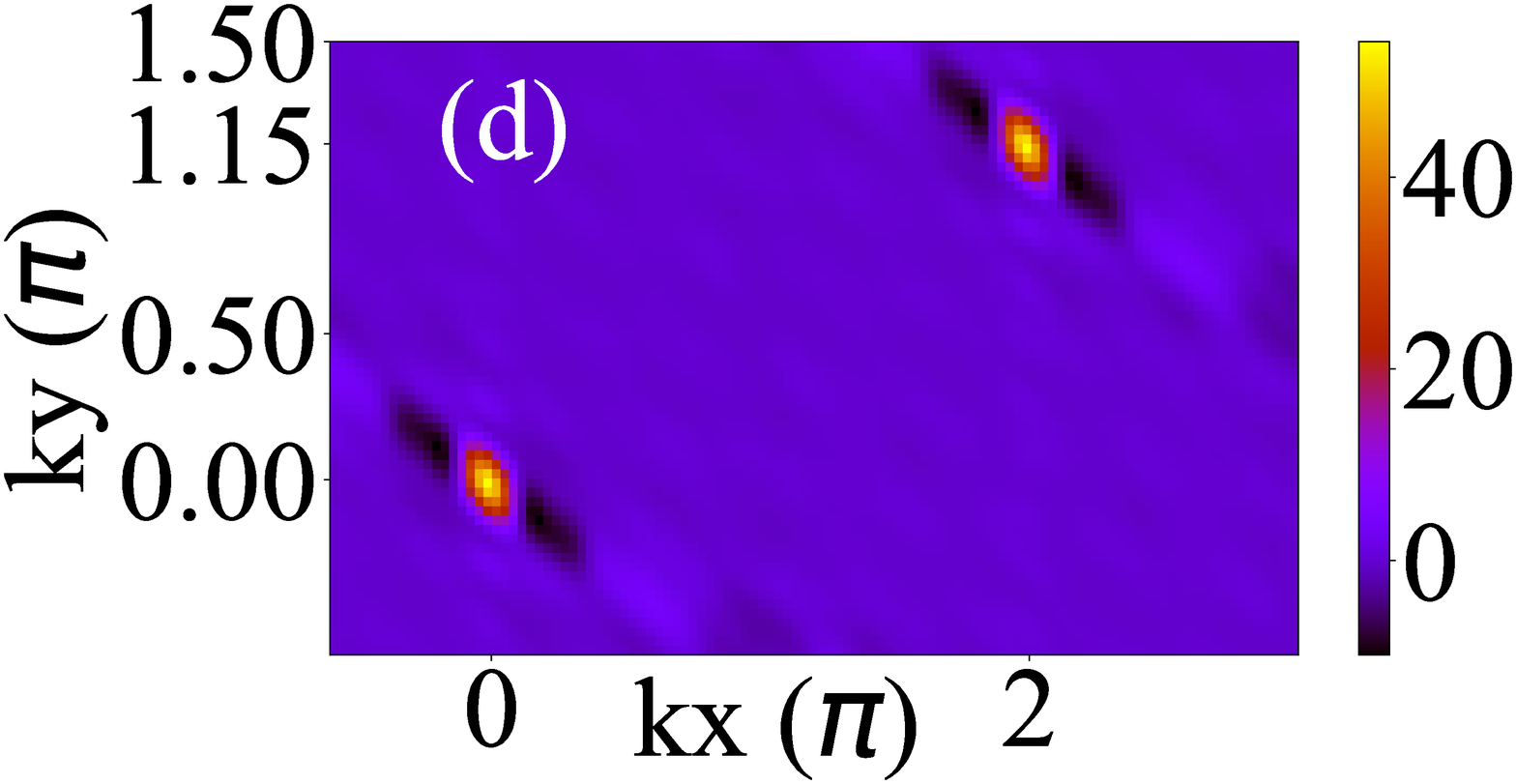}
\caption{Distribution of  $Sq\left( kx,ky\right)$
and momentum distribution $n\left( kx,ky\right)$ for the $SS$ phase  on the
triangular lattices with $L\times L\times 2$. (a) $Sq\left( kx,ky\right)$, $L$= 6
(b) $Sq\left( kx,ky\right)$, $L$= 12 (c) $n\left( kx,ky\right)$, $L$= 6 (d) $n\left( kx,ky\right)$, $L$= 12. The parameters are $\mu/U=-0.77$, $\beta%
=500$, $t/U=0.015$.}
\label{sqnx}
\end{figure}

\begin{figure}[hbt]
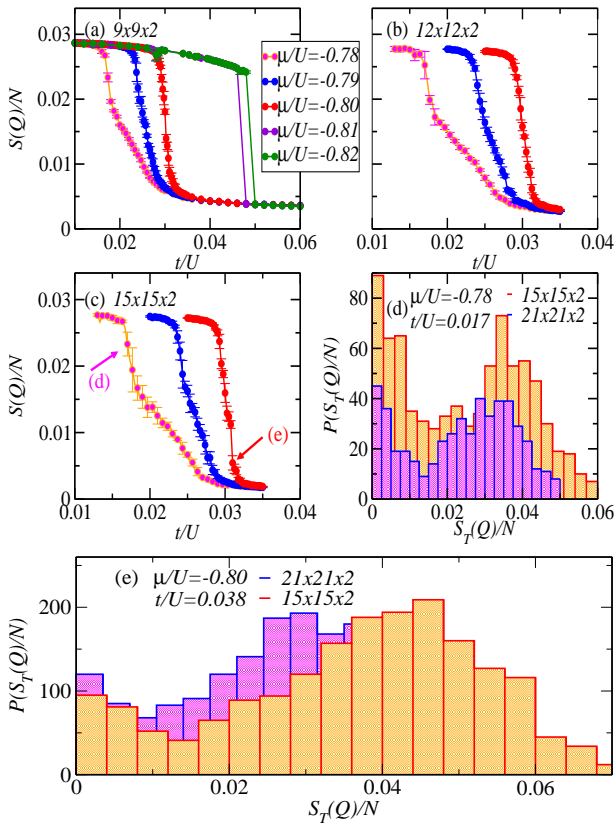

\includegraphics[width=0.45\textwidth,height=0.40\textwidth]{fig10_triangle-6-lines91215.eps}\vskip 0.1cm
\includegraphics[width=0.45\textwidth,height=0.20\textwidth]{fig10_triangle-080-sqb1-15-21}

\caption{Simulation results by the wQMC method of $S(Q)/N$ as a function of $t/U$ for the JCH
model on the triangular lattices with parameter $\protect\beta=500$, the fixed sizes 
are (a) $L=9$, (b) $L=12$ and (c) $L=15$. The colored lines denote different values of $\protect\mu/U$ near the phase. (d) Histogram of $S_T(Q)/N$ obtained at the phase transition point $\mu/U=-0.78$ and $t/U=0.017$ for the system size of $L=15$ (yellow color) and $21$ (purple color). (e) Histogram of $S_T(Q)/N$ obtained at the phase transition point $\mu/U=-0.80$ and $t/U=0.038$ for the system size of $L=15$ (yellow color) and $21$ (purple color).}
\label{sixlines}
\end{figure}

Examples for the structure factor are shown in Figs.~\ref{sqnx}(a)-(b), with $L=6$ and $12$ respectively. The two peaks of $S\left( Q\right)$ with maximum value
are located at $(0,0)$ and ($2\pi$, $1.15\pi$) because of a partially translational invariance between densities on lattices.
In the $SS$ phase at $\mu /U=-0.77$, $t/U=0.015$, we observe a second
maximum which is at $Q=( 4/3\pi, 0)$, with a $\sqrt3 \times \sqrt3$ diagonal long-range order of filling 2/3\cite{Jap,sqrt3}.
This peak indicates the presence of a density ordering in the liquid phase, 
which defines a $SS$.
The momentum distribution in Eq. (\ref{eq:nk}) is also obtained, which was observed experimentally. The two peaks of $n\left( k\right)$ with maximum value are located at $(0,0)$ and ($2\pi$, $1.15\pi$), which is the sign of the $SF$ phase, as shown in Figs.~\ref{sqnx}(c)-(d).

In the hard-core BH model on the triangular lattices, it has been verified that the types of 
$SS$-$SF$ transitions are first-order or continuous, depending on the regimes of chosen $\mu$ \cite{Jap,sqrt3}. In Fig.~\ref{trimc}, with the fixed parameter $\mu /U=-0.77$, we show the curves of $\rho$, $\rho _{s}$, and $S(Q)/N$  as a function of $t/U$ for the JCH model, and the $SS$-$SF$ phase transition is continuous, and the $SII$-$SS$ phase transition is also continuous. 

In Fig.~\ref{sixlines}, we carefully examine the phase transitions among the $SII$, $SIII$, $SS$ and $SF$ phases, with respect to the  deviation from the parameter of $\mu /U=-0.77$. Figs.~\ref{sixlines}(a), (b) and (c) show the $S(Q)/N$ as a function of $t/U$ with system sizes $L$=$9, 12$ and $15$, respectively. When $\mu /U=-0.78$, the first-order phase transition occurs between the $SII$ and the $SS$ phase. 

Simultaneously, the transition from the $SS$ to the $SF$ phases is continuous. It is still continuous when $\mu /U=-0.79$.
When $\mu /U=-0.80$, it is found that there is an obvious jump of the $S(Q)/N$ at $t/U=0.031$ (see the red line in Fig.~\ref{trimc}(c)). The bimodal distribution of total structural factors $S_T(Q)/N$ in Fig.~\ref{sixlines}(d) and Fig.~\ref{sixlines}(e) are displayed to confirm the first order (discontinuous) property of the $SS$-$SF$ phase transition.

\section{conclusion}
\label{sec:con}
In conclusion, we have systematically investigated the hard-core JCH model on one-dimensional lattices, square lattices and triangular lattices using the wQMC method. Three important quantities, the excitation densities $\rho$,
structural factors $S(Q)/N$, and superfluid stiffness $\rho_s$ are measured.

For the bipartite lattice such as the one-dimensional chain and square lattices, 
 we clarified that the previously found $SS$ phase by the 
MF method is not really stable.
By doping vacancies or excitations on the $SI$ phase, $S(Q)/N$ and $\rho _s^x$ 
jump obviously 
in the thermodynamic limit and a two-peak histogram of $S_T(Q)/N$ emerges, which represents clear first-order transitions between the $SF$ and $SI$ phases. We found that the phase boundaries between the empty-$SF$ phase transition and the $MI$-$SF$ phase transition are the same with $MF$ methods.

For the JCH model on the triangular lattices, the phase diagram obtained from the MF method is shown as background, which contains the empty, $SII$, $SIII$, $MI$, $SS$ and $SF$ phases in the $t$-$\mu$ plane. The numerical results obtained by the wQMC methods are also included to locate the regimes of the stable $SS$ phase in the thermodynamic limit. The peaks of $Sq(kx, ky)$ and the experimentally observed momentum distribution $n(kx, ky)$ confirm the presence of the $SS$ phase.

Surrounded by three different phases, the phase transitions between the $SS$ phase and other phases are more complicated and interesting. The $SIII$-$SS$ transition is confirmed to be continuous. However, the $SII$-$SS$ transition can be obviously first-order, depending on the regimes of $\mu$ chosen. The $SS$-$SF$ transition is similar to that in the hard-core BH model and this transition is first order (far away from the symmetric point) or continuous (near the symmetric point).

%
% Scaning the parameter $t/U$ from the $SIII$ phase ($\rho $=2/3) to the $SS$ phase,  both $S(Q)/N$ and $\rho _{s}$ change continuely, however, scaning the parameter $t/U$ from the $SII$ phase ($\rho $=1/3) to the $SS$ phase,  $S(Q)/N$ change strangely.

\begin{acknowledgments}
We thank Prof. N. V. Prokof for sharing his codes with WZ during the 2017 Many Electron Collaboration Summer School held at the Simons Center of Stony Brook university. 
WZ is also grateful for the invaluable discussion on simulations with Zhiyuan Yao, Lode Pollet. J. Zhang is supported by the Open Project
from the State Key Laboratory of Quantum Optics and Quantum Optics Devices
at Shanxi Province (KF201808). W. Zhang is supported by the Science Foundation for Youths of Shanxi Province under Grant No. 201901D211082.
\end{acknowledgments}

\appendix

\section{The geometries of one-dimensional chain and square lattices}

\begin{figure}[htb]
\includegraphics[width=0.3\textwidth]{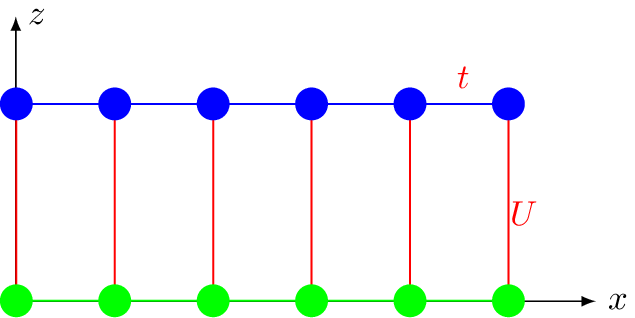} \vskip 0.1cm
\includegraphics[width=0.4\textwidth]{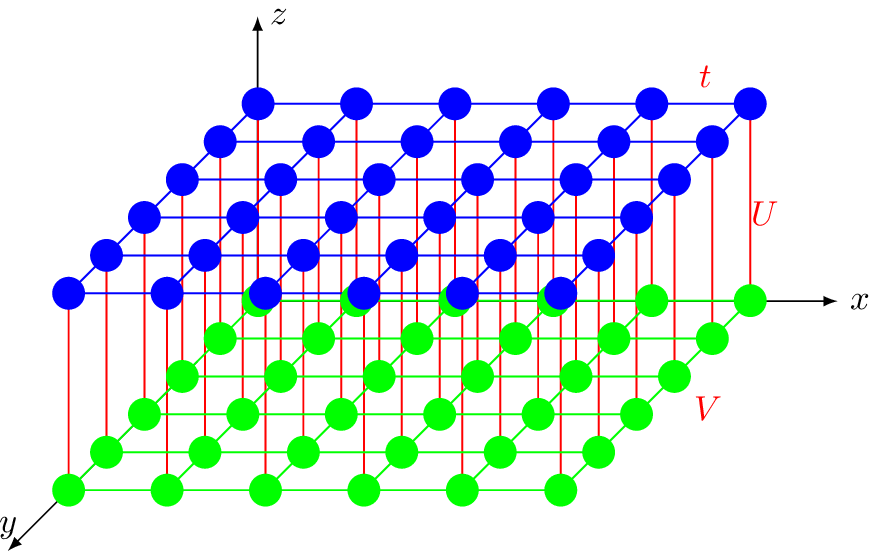} %
\caption{The mapped two-layer one-dimensional chain and square lattices from the JCH model, where the top and
bottom layers can be denoted as photon and atom layers respectively. The
hopping of photons $t$, atom-photon coupling $U$, and interactions between atoms excitations $V$ are labeled.}
\label{fig:latt1}
\end{figure}

In Fig.\ref{fig:latt1} we mapped JCH model a two-layer triangular lattice, here, we show the JCH model geometry of one dimensional chain and square lattices

\section{Comparison with exact  diagonalization}
Before the large-scale simulations, we compare the results of small systems
such as $L=4$ by the wQMC method with exact  diagonalization at parameters $t/U=0.05$ and $V/U=0.4$. We
perform QMC at the fixed temperature $\beta=200$ while exact  diagonalization with $\beta=100,~200,~300,~400$ and $500$. The quantities such as the excitation densities $\rho$, the structural factors $S(\pi)/L$, and the energy densities $E$
are shown and consistent with each other.

\begin{figure}[t]
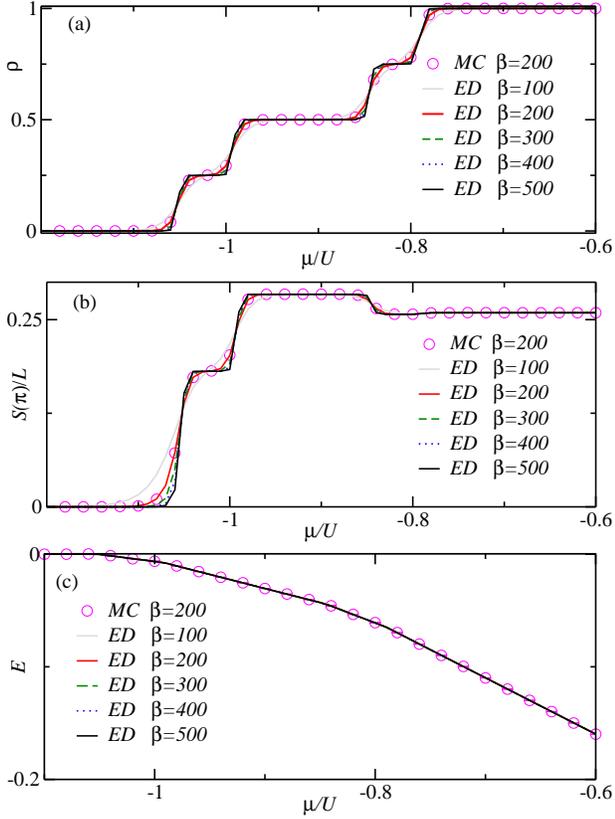

\includegraphics[width=0.45\textwidth]{fig12_a_421na.eps} \vskip 0.1cm
\includegraphics[width=0.45\textwidth]{fig12_b_421sqb.eps} %
\includegraphics[width=0.45\textwidth]{fig12_c_421Ec.eps}
\caption{Comparison of the simulation results of the one-dimensional JCH lattice
with $L=4$ by both ED and wQMC methods (a) $\protect\rho$ (b) $Sq(\protect\pi%
)/L$ (c) $E$ for various temperatures $\protect\beta=100-500$.}
\label{1ds}
\end{figure}

\section{MF methods}
\begin{figure}[b]
\includegraphics[width=0.45\textwidth]{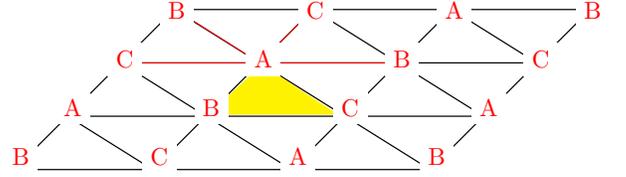} %
\caption{(Color online) A triangular lattice by three sublattices $A$, $B$ and $C$. The 
triangle denoted by the yellow color is the cluster treated exactly. The interaction and hopping 
illustrated by the red lines such as AB, AC are decoupled approximately.}
\label{fig:mft}
\end{figure}

In the CMF frame, the total Hamiltonian is taken to be
$H_{tot}=H_{mf}+H_{ed}$, where the exactly treated $H_{ed}$ is the Hamiltonian inside
the cluster, such as the triangular lattice (ABC) illustrated by a yellow color
in Fig.~\ref{fig:mft}.
The Hamiltonian $H_{mf}$ is given by
\begin{equation}
\begin{aligned}
H_{mf}=&-q t \sum_{i,j\in ce}[(a_i^\dagger+a_i) \Psi_j+(a_j^\dagger+a_j)\Psi_i-2\Psi_i \Psi_j]\\
&+q V\sum_{i,j\in ce}({n_i^\sigma \rho_j^\sigma+n_j^\sigma \rho_i^\sigma-\rho_i^\sigma \rho_j^\sigma }).
\label{g}
\end{aligned}
\end{equation}
Here, the site $A$ connects $B$ and $C$  by
the four red lines. Therefore, on average, $q$ should be equal to 2 for the 
triangular lattices.
The  symbol $ce$ means the sites along the edge of the yellow cluster. The symbol
$\Psi_i=\la a_i\ra$ is the superfluid order parameter, and
$\rho_i^\sigma=\la n_i^\sigma\ra$ is the number of atomic excitations.

%
%%%
%By systematically scaling the cluster size $nc$, one may obtain quantitative exact results for 2D-systems when comparing to more extensive QMC studies~\cite{Dynamical critical, zhaojize, Singh2018}. Therefore, we consider different cluster sizes and tessellations of the triangular lattice to examine the stability of our findings. In Fig.~\ref{fig:cmf_graphs}, we exemplify different cluster sizes employed in the study.

In practice, we determine the self-consistent solutions of $\rho_i^\sigma$ and $\Psi_i$ by iterative calculation of the ground state of the cluster-system until the mean fields have converged.
% While for small cluster sizes such as Fig.~\ref{fig:cmf_graphs}(E) the ground state of the cluster may be obtained by exact-diagonalization techniques, for the larger cluster sizes we employ the DMRG simulation. Following Ref.~\cite{clusterDMRG} in this MF scheme already a low number of matrix-states is sufficient. In the present examples we choose $30$ and $60$ states for $nc=12$ site cluster results shown below.
In Figs.~\ref{hdph} and \ref{xt}, $\Psi$ is plotted as background for the phase diagram.
%
%The solid or density wave orders denoted by  $\Delta\rho^a$, $\Delta\rho^{\sigma}$, and $\Delta \Psi$ are defined by
%\be \label{eq:2}
%\delta A  =\frac{1}{nc}\sum_{i\in c} |A_i-\bar{A}|,~ \bar{A}=\frac{1}{nc}\sum_{i\in c}\bar A_i.
%\ee
%We also define the total excitation $\rho=\rho^a+\rho^\sigma$~\cite{SMI}, and  $\Delta \rho = \frac{1}{2}({\Delta\rho^a+\Delta\rho^\sigma})$.

\end{document}